\input{aipcheck}

\documentclass[
  ,final            
  ,draft            
  ,numberedheadings 
  ]
  {aipproc}

\layoutstyle{6x9}

\usepackage{bm}
\usepackage{amsmath}
\usepackage{amssymb}

\def\LamFOF{\Lambda \text{(1405)}}

\def\KbarN{\bar{K} N}

\def\Rho{\text{P}}
\def\PE{\Rho _{\text{E}}}

\def\PB{\text{P}_{\text{B}}}
\def\PS{\text{P}_{\text{S}}}

\def\fm{~\text{fm}}

\begin{document}

\title{Probing internal structure of $\LamFOF$ in 
meson-baryon dynamics with chiral symmetry}

\classification{
13.75.Jz, 
14.20-c,  
11.30.Rd  
}
\keywords{$\Lambda \text{(1405)}$ structure, 
  meson-baryon chiral dynamics, 
current couplings to resonances}

\author{Takayasu Sekihara}{
  address={Department of Physics, Kyoto University, Kyoto 606-8502, Japan}
  ,altaddress={Yukawa Institute for Theoretical Physics, 
    Kyoto University, Kyoto 606-8502, Japan} 
}

\author{Tetsuo Hyodo}{
  address={Department of Physics, Tokyo Institute of Technology, 
  Tokyo 152-8551, Japan}
}

\author{Daisuke Jido}{
  address={Yukawa Institute for Theoretical Physics, 
    Kyoto University, Kyoto 606-8502, Japan}
}

\begin{abstract}
  The internal structure of the resonant $\LamFOF$ state is
  investigated based on meson-baryon coupled-channels chiral dynamics,
  by evaluating density distributions obtained from the form factors of
  the $\LamFOF$ state.  
  The form factors are extracted from current-coupled scattering
  amplitudes in which the current is coupled to the constituent
  hadrons inside $\LamFOF$.
%
  Using several probe interactions and channel decomposition, we find
  that the resonant $\LamFOF$ state is dominantly composed of widely
  spread $\bar{K}$ around $N$, with a small fraction of the escaping
  $\pi \Sigma$ component.
\end{abstract}

\maketitle


\section{Introduction}

It is one of the important issues in hadron physics to clarify
properties of the resonant $\LamFOF$ state, which is an excited baryon
with spin-parity $J^{P}=1/2^{-}$, isospin $I=0$, and strangeness
$S=-1$, and is located just below the threshold of antikaon
($\bar{K}$) and nucleon ($N$).  The $\LamFOF$ state has been
considered as a $\KbarN$ quasi-bound state~\cite{Dalitz:1960du}.
Recent theoretical studies have also suggested that the $\LamFOF$ is
dynamically generated in meson-baryon coupled-channels chiral
dynamics, or so-called chiral unitary approach~\cite{Kaiser:1995eg,
  Oset:1997it,Oller:2000fj,Lutz:2001yb,Jido:2003cb},
reproducing well the low-energy $K^{-}p$ cross sections as well as the
$\LamFOF$ peak in $\pi \Sigma$ mass spectrum.  Moreover, the chiral
unitary approach predicts double-pole structure for
$\LamFOF$~\cite{Jido:2003cb} and one of the poles is expected to 
originate from the $\KbarN$ bound
state~\cite{Hyodo:2007jq,Hyodo:2008xr}.  Some approaches for the
survey on the $\LamFOF$ structure in experiments have been proposed,
{\it e.g.}, in Refs.~\cite{Jido:2009jf,Cho:2010db}.

If $\LamFOF$ is dominated by the $\KbarN$ quasibound state with a
small binding energy, one can expect that $\LamFOF$ has a larger size
than typical ground state baryons dominated by genuine $qqq$ components.
Motivated by this expectation, in Ref.~\cite{Sekihara:2008qk} 
we investigate the internal structure of the resonant $\LamFOF$ state
by evaluating density distributions obtained from the form factors on
the $\LamFOF$ pole originating from the $\KbarN$ bound state.
In our study we extract the form factors of $\LamFOF$ directly from the
current-coupled scattering amplitude, which involves a response of
$\LamFOF$ to the external current.  The current-coupled scattering
amplitude is evaluated in a microscopic way, that is, by considering 
current couplings to the constituent hadrons inside $\LamFOF$.  
The wave functions and form factors of $\LamFOF$ were studied also in
Ref.~\cite{YamagataSekihara:2010pj} in a cut-off scheme within chiral
unitary approach.  Their results would not show much difference with
respect to our results, except for the high momentum region compared
to the cut-off scale introduced in the cut-off scheme.

\section{Internal structure of  $\bm{\Lambda \text{(1405)}}$}

In chiral unitary approach, $\LamFOF$ is dynamically generated in the
meson-baryon coupled-channels amplitude $T_{ij}$ with the channel
indices $i$ and $j$.  In order to observe response of the $\LamFOF$
state with respect to the conserved probe current, we evaluate
current-coupled scattering amplitude $T_{\gamma ij}^{\mu}$ in a charge
conserved way, considering current couplings to the constituent
hadrons~\cite{Borasoy:2005zg}.  Then the form factor,
$F^{\mu}(Q^{2})$, can be extracted
by~\cite{Sekihara:2008qk,Jido:2002yz},
\begin{equation}
F^{\mu} (Q^{2}) 
=  
- \frac{(z^{\prime}-Z_{\text{R}})T_{\gamma ij}^{\mu}
(z^{\prime}, \, z; \, Q^{2} )}{T_{ij} (z)} 
\Bigg |_{z \to Z_{\text{R}}}
\Bigg |_{z^{\prime} \to Z_{\text{R}}} , 
\label{eq:Res_scheme}
\end{equation}
where $Q^{2}$ is squared current momentum, $z^{(\prime )}$
center-of-mass energy in complex plane, and $Z_{\text{R}}$ the
$\LamFOF$ pole position.  The details of the
calculation 
are given in Ref.~\cite{Sekihara:2008qk}.

Now let us show numerical results for the internal structure of the
resonant $\LamFOF$.  First, in order to pin down the dominant
component of the $\LamFOF$ structure we evaluate the values of the
baryonic and strangeness form factors at $Q^{2}=0$ with various
components of $\LamFOF$, which correspond to individual contributions
from meson-baryon channels and contact term to the baryon number and
strangeness of the system.  As a result, we find that the $\KbarN
(I=0)$ channel generates $0.994 + 0.048i$ of the baryonic and
(opposite sign of) strangeness charges of $\LamFOF$, which is unity,
whereas the other components such as $\pi\Sigma (I=0)$ channel are
negligibly small~\cite{Sekihara:2008qk}.  
This result indicates that the $\KbarN (I=0)$
channel dominates the total baryonic and strangeness charges of
$\LamFOF$, giving more than $99 \%$ of the charges.  The $\KbarN$
channel dominance for the $\LamFOF$ structure is caused by the large
coupling strength of $\LamFOF$ to $\KbarN$, $g_{\KbarN}$.

\begin{figure}[!t]
  \centering
  \begin{tabular*}{\textwidth}{@{\extracolsep{\fill}}cc}
    \includegraphics[width=0.485\textwidth]{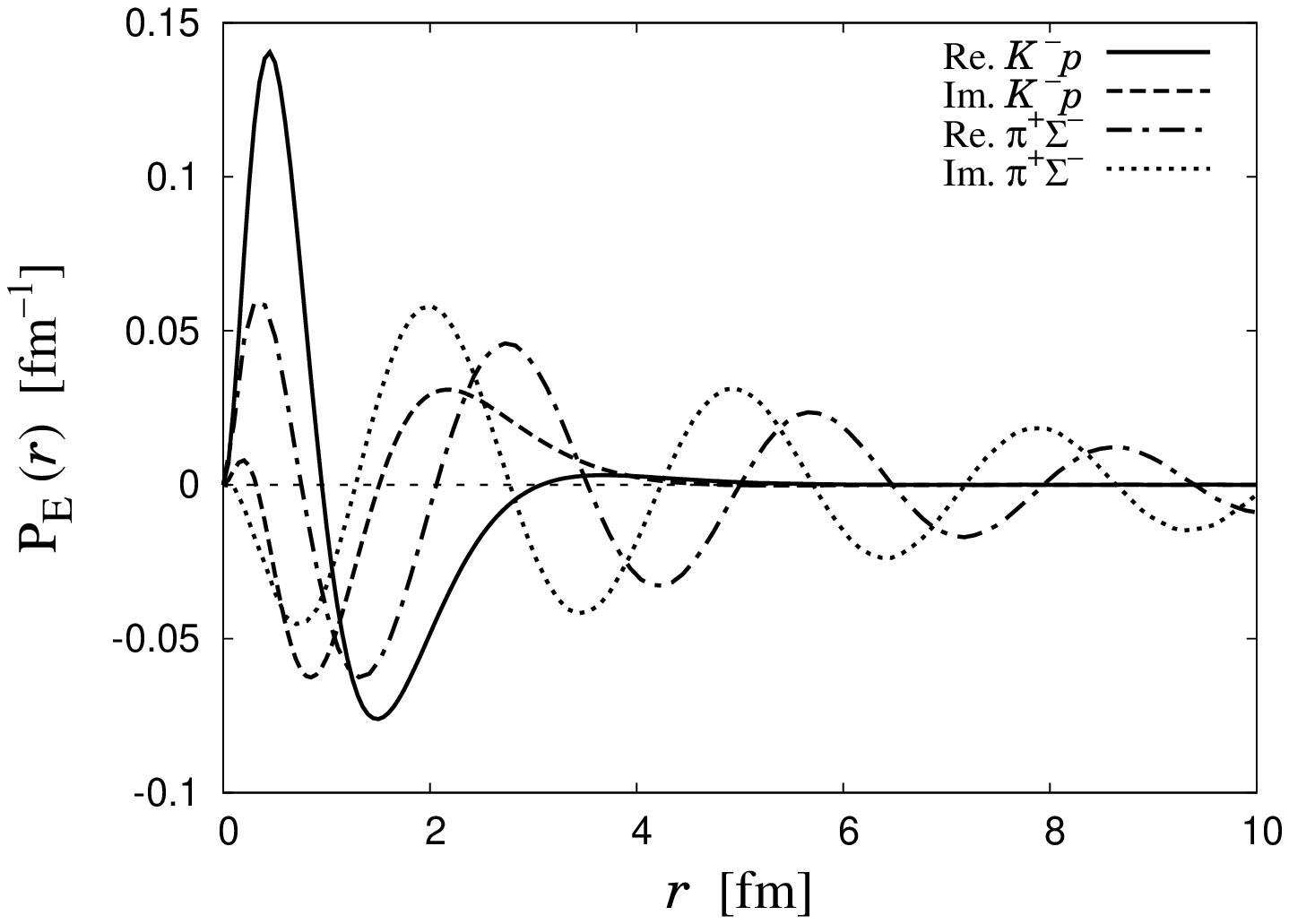} &
    \includegraphics[width=0.485\textwidth]{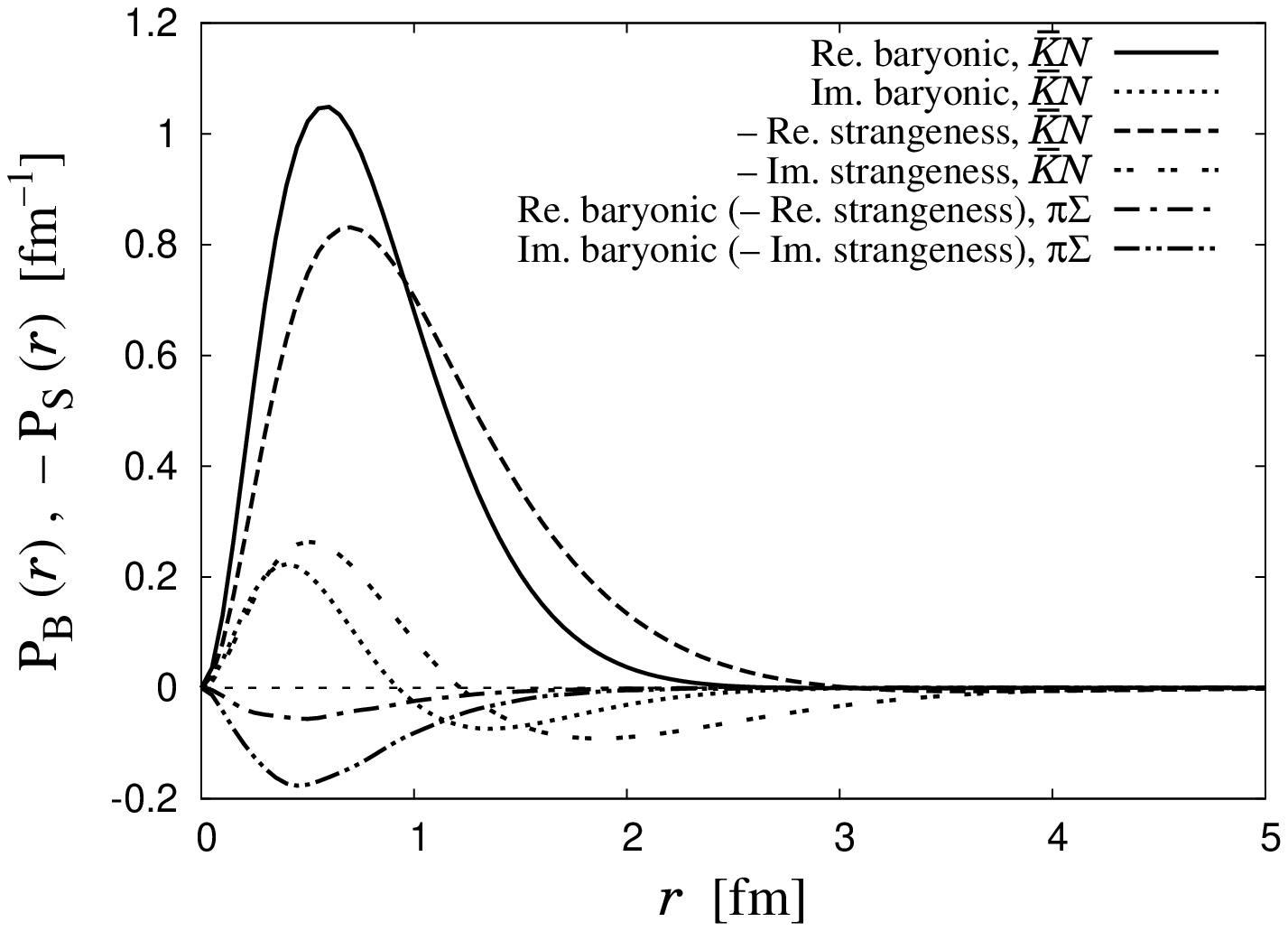} \\
  \end{tabular*}
  \caption{
    Electric ($\PE$, left), baryonic and strangeness ($\PB$ and $\PS$,
    right) density distributions of $\LamFOF$ in each component.  Here
    $\PE$ is shown in particle basis whereas $\PB$ and $\PS$ are in
    isospin basis~\cite{Sekihara:2008qk}. 
    \vspace{-10pt}
  }
  \label{fig:Rho_pole}
\end{figure}

Next we show the meson-baryon components of the electric, baryonic,
and strangeness density distributions ($\PE$, $\PB$, and $\PS$,
respectively) in Fig.~\ref{fig:Rho_pole}.  Here $\PS$ is presented
with opposite sign for comparison.
From $\PE$, we can see that the negative (positive) charge
distribution appears in $\LamFOF$ due to the 
existence of lighter $K^{-}$ (heavier
$p$) in the outside (inside) region, bearing in mind the
$\KbarN$ dominance for $\LamFOF$.  Also it is interesting to see the
dumping oscillation in $\pi ^{+} \Sigma ^{-}$ (equivalently $\pi^{-}
\Sigma^{+}$ with the opposite sign) component in $\PE$, which can be
interpreted as the decay of the system into the $\pi\Sigma$ channels
through the photon coupling to the intermediate $\pi$, as discussed in
Ref.~\cite{Sekihara:2008qk}.  This oscillation behavior is, however,
not observed in the total electric density distribution due to the
cancellation of $\pi ^{+} \Sigma ^{-}$ and $\pi ^{-} \Sigma ^{+}$
components.
From $\PB$ and $\PS$, on the other hand, we can separately observe the
$\bar{K}$ and $N$ components inside $\LamFOF$, because the baryonic
(strangeness) current probes the $N$ ($\bar{K}$) distribution.  These
distributions indicate that inside $\LamFOF$ the $\bar{K}$ component
has longer tail than the $N$ component and $\bar{K}$ distribution
largely exceeds typical hadronic size $\lesssim 1 \fm$.

\section{Summary}

We have investigated the internal structure of the resonant $\LamFOF$
state in the chiral unitary approach, in which $\LamFOF$ is
dynamically generated in meson-baryon coupled-channels chiral
dynamics.  Making conserved probe current couple to $\LamFOF$ in a
charge conserved way, we have observed that $\KbarN$ component gives
more than $99 \%$ of the total baryon number and strangeness of
$\LamFOF$.  The electric density distribution has shown that in
$\LamFOF$ lighter $K^{-}$ (heavier $p$) exists in the outside (inside)
region and the escaping $\pi \Sigma$ component appears.  Also from the
baryonic and strangeness density distributions we have found that
inside $\LamFOF$ the $\bar{K}$ component has longer tail than the $N$
component and $\bar{K}$ distribution largely exceeds typical hadronic
size $\lesssim 1 \fm$.


\begin{theacknowledgments}
  This work is partly supported by the Grand-in-Aid for Scientific
  Research from MEXT and JSPS (No.  21840026, 
  22105507, 
  22740161, 
  and 22-3389
  ). 

\end{theacknowledgments}



\bibliographystyle{aipproc}   


\IfFileExists{\jobname.bbl}{}
 {\typeout{}
  \typeout{******************************************}
  \typeout{** Please run "bibtex \jobname" to optain}
  \typeout{** the bibliography and then re-run LaTeX}
  \typeout{** twice to fix the references!}
  \typeout{******************************************}
  \typeout{}
 }

\end{document}